# From stand-up to start-up: exploring entrepreneurship competences and STEM women's intention

Authors Armuña, Cristina. Universidad Politécnica de Madrid, Spain

Ramos, Sergio. Universidad Nacional de Educación a Distancia (UNED), Spain

Juan, Jesús. Universidad Politécnica de Madrid, Spain

Feijoo, Claudio. Universidad Politécnica de Madrid, Spain

Arenal, Alberto. Universidad Politécnica de Madrid, Spain

**Abstract**

This paper contributes to the study of the entrepreneurship competences debate. Taking as reference the Entrepreneurship Competences Framework proposed by the EC as a common guide to inspire entrepreneurship education, we examine a paradigmatic group of potential STEM entrepreneurs regarding their perception of competences and the link with entrepreneurial intentions. Slight differences are obtained between women and men, only significant on the *recognition of the potential value of an idea*, the *digital know-how*, the *financial and economic know-how* and the *team working skills*. In a next step, factor analysis provides a different -but meaningful- clustering of competences from those in the theoretical framework and resulting in a better model fit. This new clustering comprises *opportunity and ideas*, *commitment*, *decision making, organizing and specific knowledge.* In a final step, competences about *opportunity and ideas*, *commitment*, *decision making* and *organizing* groups prove to be more related to higher entrepreneurship intention. Gender do not have a moderating factor on this relationship. As a final result of interest from the paper is that the entrepreneurship intentions of this STEM group do not validate the usual assumption that women have fewer entrepreneurship intentions than men.

**KEYWORDS:** Entrepreneurship intention, women entrepreneurship intention, entrepreneurship competences, entrepreneurship education, gender in entrepreneurship

## INTRODUCTION

Entrepreneurship has become of primary relevance in the current economic context, being one of the key priorities for policy plans and initiatives as a means to foster economic and social progress (Arenal et al., 2017), especially in sectors related to the STEM disciplines because of their impact on innovation and economic growth (O'Shea, 2007). However, these efforts contrast with a relatively low entrepreneurship intention rate in some geographies (GEM[1])[2] and the commonly recognised gender gap regarding entrepreneurship intention (Guerrero et al., 2016; Kelley et al., 2016; Santos, Roomi, and Liñan, 2016) is particularly relevant on these areas (Sánchez Cañizares and Fuentes García, 2010; Dilli and Westerhuis, 2018).

In this context, entrepreneurship education is an essential driver that should be promoted along the entire education lifecycle, from primary and secondary school, to higher education and beyond as a mean to foster entrepreneurship (World Economic Forum, 2009). As the intention to do something is considered the best single predictor of effectively doing it (Krueger and Brazeal, 1994), the impact that education can have on entrepreneurship intention will sum into the policy aspiration to increase effective entrepreneurs (Schlaegel and Koenig, 2014).

---

[1] http://ec.europa.eu/social/main.jsp?catId=1223

[2] In the European Union only 10.8% of 16-64 years old population intend to start a business compared to 14.3% in North-America, 26.2% in Asia and Oceania, 32,3% in Latin-America and 33.4% in Africa (GEM, 2018). Particularising the analysis for Spain, global indexes show that entrepreneurship intention in 2017 had the third lowest rate (5.6%) (GEM, 2018) from a list of 54 countries.

Most studies exploring factors that influence entrepreneurship intentions have mainly emphasised norms, attitudes and traits (Micaela et al., 2014; Lortie and Castogiovanni, 2015). While these have presented relevant insights into how differences in intentions arise between individuals and how education can affect these intentions (Liñán, Rodríguez-Cohard and Rueda-Cantuche, 2011; Entrialgo and Iglesias, 2016), the main output of education is competence development and, although research notes that initiating a business is affected by the fostering of skills and knowledge (Sánchez, 2013; Kelley, Singer and Herrington, 2016; Santos, Marques and Ferreira, 2018), the relationship between competences and entrepreneurship intention has gained much less attention.

In this sense, although many scholars have explored the definition, design and purpose of entrepreneurship education programs (EEP) since the 1980s (Bechard and Gregoire, 2005), including attempts to identify the key entrepreneurship competences for higher education organisations in order to help develop them (Robles and Zárraga-Rodríguez, 2015), there is still debate about the conception of the abilities that should be taught (Bacigalupo et al., 2016).

This is of particular interest from the gender perspective. When exploring the relationship between the perception of barriers and entrepreneurship intentions, research has found a lack of competence perceived to be more important for women than men, in the sense of having the necessary knowledge, skills, and experience to be an entrepreneur (Shinnar, Giacomin and Janssen, 2012).The understanding of competences perceived by female entrepreneurs and the relationship of their acquisition to the different venture stages are understudied topics, despite its importance when designing education and training initiatives (Mitchelmore and Rowley, 2013).

Women usually exhibit lower rates of entrepreneurship intention than men (Novak et al., 2012; Guerrero et al., 2016; Kelley et al., 2016; Santos, Roomi and Linan, 2016), particularly in highly knowledge-intensive business sectors (Dilli and Westerhuis, 2018), and education can play a determinant role in promoting competences that affect the entrepreneurial intentions of both male and female students. Studying the competence-intention relationship and the potential effect of gender, can contribute to shedding light on how EEPs should be shaped and adapted for specific needs.

As a contribution to this debate, we thus attempt to deepen in the exploration and understanding of entrepreneurship competences and their connection to entrepreneurial intention in a paradigmatic cohort of STEM profile individuals who have shown interest in joining an entrepreneurship education and incubation program, the ActuaUPM[3] program of the Technical University of Madrid (UPM). Assuming that they hold entrepreneurship intentions, the sample is of especial interest because they represent potential STEM entrepreneurs. The objective of this paper is not to test how a particular EEP impacts competences, but to examine this group of STEM potential entrepreneurs regarding their perception of competences and explore which of those competences are related to a higher or lower entrepreneurship intention, considering gender as a moderating factor.

---

[3] The ActuaUPM entrepreneurship education initiative is aimed at students (31.7% of woman at UPM in 2015), teachers and researchers (24.7% of woman at UPM in 2015) of the Technical University of Madrid (UPM), and is structured in three phases through which participants evolve, from the idea stage to business plan development and, in some cases, to the formal consolidation of a new venture. The program has already run on 14 occasions, and a total of 146 firms were founded from UPM during the period 1998 - 2013, most linked to ActuaUPM.

Taking the Entrepreneurship Competences Framework (EntrComp), proposed by the European Commission (EC) as a common reference with which to define EEPs (Bacigalupo et al., 2016), our results contribute to the identification of competences whose promotion could increase STEM entrepreneurship mindset. The profile of the sample can be characterised according to their self-perception of competences, noting differences between women and men, and exploring which are more closely related to entrepreneurship intentions, considering the potential gender moderating effect.

**THEORETICAL FRAMEWORK AND HYPOTHESIS**

**Entrepreneurship competences and intentions**

Competences are understood as the set of skills and knowledge that a person achieves through education or experience (Burgoyne, 1993; Parry, 1998; Man, Lau and Chan, 2002; Cheng, Dainty and Moore, 2003; Bacigalupo *et al.*, 2016). They are considered personal attributes that affect the decisions we take and the way we put them into action and, consequently, they can foster or decrease our intentions to do something (Spencer Jr. and Spencer, 1993). Among the mechanism of personal agency, the Social Cognitive Theory point the relevance of people's beliefs about their capabilities -*self-efficacy perception*- to exercise control over events that affect their lives is considered determinant (Bandura, 1989), being the actions more affected by what people believe they can do than with objective facts (Markman, Baron and Balkin, 2005).

The concept of entrepreneurship competences has acquired a great relevance in recent decades because of the importance that entrepreneurship has gained as an economic driver in society, and due to the conception of entrepreneurs as individuals with the capacity to create value, usually linked to the creation and development of businesses but desirable for society in general (IEG World Bank, 2013; Council of the European Union, 2018). In this sense, entrepreneurship intentions, their configuration and what influences them, have been extensively explored in research, although basically linked to three precedents: a positive or negative attitude towards a career as an entrepreneur, the perceived capability to start a business and the perception of the social acceptance (Lortie and Castogiovanni, 2015; Micaela et al., 2014).

Nevertheless, competences also play a role in entrepreneurship intentions and education is considered an enabler (Kelley et al., 2016; Sánchez, 2013). Their presence on policy agendas, and specifically via entrepreneurship education, has been increased as a facilitator of these entrepreneurship competences, that can be trained and developed (Burgoyne, 1993; Parry, 1998). There is still, however, debate about what these entrepreneurship competences are, and which are more related to the intention to start a business (Bacigalupo *et al.*, 2016).

Although entrepreneurship competences have been usually linked to management skills it is commonly accepted that entrepreneurship activities involve a wider range of competences beyond the management of businesses (Bird, 1995; Man, Lau and Chan, 2002; Lerner and Almor, 2003). Entrepreneurship education has evolved following this approach, from where it was initially not considered a specific discipline that required a different methodology to that of standard business and management programs (Jones and English, 2001), now addresses an alternative path through innovative teaching methods, contents and scope, particularly promoting abilities rather than specific management knowledge (Vesper and Gartner, 1997; Rae, 2005; Haase and Lautenschläger, 2011; Boyles and College, 2012; Thrane *et al.*, 2016). Although most empirical studies in which authors make pre- and post-treatment note that entrepreneurship education programs have a positive impact on the entrepreneurship intentions of students (Souitaris, Zerbinati and Al-Laham, 2007; Mueller, 2011; Sánchez, 2013), some others see a negative effect (Oosterbeek, van Praag and Ijsselstein, 2010; Patricia and Silangen, 2016), particularly on female university students (Westhead and Solesvik, 2004). Derived from this gap, to know more about the competences that people who hold entrepreneurship intention perceive, the potential differences by gender and the relationship with a higher or lower intention emerge as a field of interest with application to education initiatives and it is the focus of this research.

One of the main challenges in research has been to group these entrepreneurship competences into comprehensive categories in order to provide further understanding and insights. Bird (1995) defined entrepreneurial competencies as *"underlying characteristics such as generic and specific knowledge, motives, traits, self-images, social roles, and skills which result in venture birth, survival, and/or growth"*. Man, et al. (2002) suggested that *"entrepreneurial competencies are considered a higher-level characteristic encompassing personality traits, skills and knowledge, and therefore can be seen as the total ability of the entrepreneur to perform a job role successfully"* and reviewed research into partial competences in order to propose a framework including six categories: *opportunity competences*, including those related to the recognition opportunities in the market, *relationship competences*, referring to interaction with people, *conceptual competences*, such as risk taking or being proactive, *organising competences*, such as team building or leadership, *strategic competences*, including project management, and *commitment competences*, including the ability to work hard.

Mitchelmore and Rowley (2010; 2013) also proposed a conceptual framework, using a literature review to create four categories: *entrepreneurial competencies*, *business and management competencies*, *human relations competencies* and *conceptual and relationship competencies*, reporting the difficulties to narrow the definitions of competences, competencies, skills or abilities, the separation between personal traits and knowledge acquired and the needs to deep into specific research involving women.

The self-perception of entrepreneurship competences and their relationship towards intention has been also addressed under the concept of *self-efficacy*, understood as people's beliefs about their capabilities to produce effects which operate action through motivational, cognitive, and affective intervening processes (Bandura, 1989, 1994). In this sense the *Entrepreneurial self-efficacy construct (ESE) proposed by* Chen (1998) include the self-perception of 22 competences grouped into five categories: *marketing, innovation, management, risk-taking, and financial control.* Based on Mueller (2001), Fernández-Pérez *et al.*, (2019) defines self-efficacy with five self-perceived capabilities related to *idea definition*, *business plan*, *negotiation*, *opportunity recognition* and *financial relationship* and, considering *emotional competences* as antecedents of entrepreneurship attitudes and self-efficacy, include a framework of five additional dimensions: *self-awareness*, *self-regulation*, *motivation*, *empathy* and *social skills*.

The debate is still open and, focused on entrepreneurship competences as capabilities or abilities that can be trained to create learning outcomes, the *EntrComp* model within the European Commission's strategy for entrepreneurship (Bacigalupo *et al.*, 2016) proposes a set of 15 competences that all together address the capacity to turn ideas into action. The framework gathers previous research about entrepreneurship competences, including a comprehensive review of the academic and grey literature, an in-depth analysis of case studies, desk research and a set of iterative multi-stakeholder consultations.

The interest on this framework emerges because it has been developed as a potential common reference with which to shape all types of education and training programs in order to provide citizens with the competences that are considered potentially applicable to all spheres of life, highlighting three core aspects connected to personal and professional development, including the creation of new businesses. The categorisation proposed is innovative because abilities are included according to the phases that have been identified in the process of taking ideas to reality: *identification of ideas*, *resources management* and *into action*. The conceptualisation is based on a theoretical approach which the authors suggest that over time will be further elaborated and refined to address the particular needs of specific target groups. We contribute to adding a quantitative dimension by proposing a set of variables that allow the perceptions of the set of competences in a particular group of potential STEM entrepreneurs to be measured, exploring whether the theoretical grouping is statistically supported and selecting those competences that result more linked to entrepreneurship intention. Characterized as self-perceived competences related to orient entrepreneurship intention, the instrument can be considered a proposal for *self-efficacy* scale.

The following paragraphs describe the competences included in the framework in detail and consider previous research about their relationship with entrepreneurship intention to define our hypotheses.

The first area of the EntreComp model is "Ideas and Opportunities", which reflects the capacity to recognise opportunity and the valuation of ideas to meet the challenges. This block of competences includes: (1) *spotting opportunities*, defined as the ability to find opportunities to generate value for others, recognising opportunities that need to be met and responding to challenges; (2) *creativity*, understood as the ability to develop multiple ideas that create value for others and transform these ideas into solutions; (3) *vision*, referring to the capacity to imagine a desirable future and an inspiring vision to guide strategic decision-making; (4) *valuing ideas*, representing the ability to understand and appreciate the value of ideas and their possible different types of use, developing strategies to make the most of the value generated by ideas; and (5) *ethical and sustainable thinking*, as the ability to recognise the impact of choices and behaviours, within both the community and the environment, driven by ethics and sustainability, when making decisions. The research literature notes that, in the case of women, a more unfavourable perception of the environment affect more towards a lower entrepreneurship intention (Langowitz and Minniti, 2007; Díaz-García and Jiménez-Moreno, 2010) and more pessimistic perception of business opportunities (Díaz-Casero, Sanchez and Postigo, 2002; Jack and Anderson, 2002; Camelo-Ordaz, Diánez-González and Ruiz-Navarro, 2016).

The second block in the *EntrComp* model is called "Resources" and involves the abilities to follow-up and opportunity that has been identified. It includes material and non-material attributes such as: (6) *self-awareness* and *self-efficacy*, defined as the capacity to trust our own ability to generate value for others, making the most of their strengths and weaknesses and compensating for the latter by teaming up with others; (7) *motivation and perseverance*, which refers the capacity to guide efforts through passion and perseverance to achieve goals; (8) *mobilising resources*, defined as the ability to find and use resources and knowledge responsibly, and to manage them; (9) *financial and economic literacy*, which refers to the capacity to draw up a budget as a first step, with progression towards the exploration of financial options and economic sustainability; and (10) *mobilising others*, defined as the capacity to communicate ideas clearly and with enthusiasm (communication skills) and to persuade, inspire and 'get others on board' (leadership skills). Self-efficacy is widely considered to effect the intention to become an entrepreneur (Liñán and Fayolle, 2015; Lortie and Castogiovanni, 2015). Its effect on entrepreneurship intention seems to be stronger for women (Wilson, Kickul and Marlino, 2007), in the sense that the perception of having the abilities considered necessary to run a business is more determinant to orient an entrepreneurial career. Other empirical analysis, however, conclude that entrepreneurial intentions and self-efficacy are not moderated by gender (Campo, 2011). The perception of self-efficiency, linked to the management of resources, is considered a factor of influence (Díaz-García and Jiménez-Moreno, 2010; Camelo-Ordaz, Diánez-González and Ruiz-Navarro, 2016) and the fear of failure, related to perseverance, is noted as one of the main factors explaining lower intentions in women than in men (Camelo-Ordaz, Diánez-González and Ruiz-Navarro, 2016).

The third block of the framework is the "Into Action" competences, reflecting features regarding the ability to transform ideas into reality and including: (11) *taking the initiative*, defined as the ability to initiate value-creating activities; (12) *planning and management*, referring to the ability to define a goal, create action plans and identify priorities; (13) *coping with uncertainty, ambiguity and risk*, including the perspective of not being afraid of mistakes and making choices despite uncertainty; (14) *working with others*, which involves the capacity to work together in a team based on the needs of a value-creating activity and the ability to solve conflicts and face competition positively when necessary; and (15) *learning through experience*, defined as the ability to recognise lessons learnt when taking part in value-creating activities, judging achievements and failures and learning from them. Action-oriented competences have been considered as key necessary abilities in previous models reported in the literature (Boyles and College, 2012).

In general, previous studies working on the identification of competences that are relevant for effective entrepreneurship offer a wide range of insights, but usually explore them in an isolated or partial way. For instance, starting from a list of twenty competences commonly considered key in the literature, Robles and Zárraga-Rodríguez (2015) identify strong effects from these nine: risk assumption, initiative, responsibility, dynamism, troubleshooting, search and analysis of information, results orientation, change management and quality of work. There is no consensus about other competences frequently considered: social networks development, self-control and social mobility (Robles and Zárraga-Rodríguez, 2015). When dividing the competences into 'soft' and 'hard', some studies find a significant effect from initiative, self-confidence and assertiveness (considered 'soft' competences) and human resources competence and production competence (considered 'hard') (Riyanti, Sandroto and D.W, 2016). Other analysis concludes that the ability to recognise income-generating opportunities, entrepreneurial training and skills, innovativeness, and information-seeking competencies have a significant effect on entrepreneurial intention (Al Mamun *et al.*, 2016).

Consequently, although there is no consensus about the effect of perceived competences, previous studies of potential women entrepreneurs conclude in general that perceived capabilities are the most important predictor of their entrepreneurial intention (Sen, Yilmaz and Ari, 2018), resulting the lack of competence being more important for women than men, in the sense of having the necessary knowledge, skills, and experience to be an entrepreneur (Shinnar, Giacomin and Janssen, 2012).

There are few studies of entrepreneurship intentions that include samples of women in STEM (Morton, Huang-Saad and Libarkin, 2016). Sánchez Cañizares and Fuentes García (2010) found that the gap in entrepreneurship intention is higher in the STEM fields than in social sciences, law or business majors. In this study, female students give a greater relevance as obstacles to set up a business their lack of business know how, the fear of failure and ridicule, as well as doubts regarding their own entrepreneurial capacity. Personal characteristics obtained as positively related to entrepreneurship intentions are moderated by gender, being independence at work, perseverance, creativity and tackling difficulties very relevant to women, addressing new challenges and take moderating risks are more relevant to men, and having initiative is relevant to both.

The paper therefore tests the following hypothesis in a specific STEM sample:

*H1: Female entrepreneurship competences perception is lower than men*

*H2: There is a positive relationship between competences and entrepreneurship intention*

*H3: Gender moderates the positive relationship between competences and entrepreneurial intentions so that the relationship is stronger for female STEM students compared to male STEM students*

**METHOD**

**Data collection and description of the sample**

A questionnaire-based survey was conducted in order to collect data. The questionnaire was launched in June 2017 among the participants of the ActuaUPM entrepreneurship program, which includes a contest for entrepreneurial ideas followed by a training program for selected teams. The questionnaire was structured as a voluntary online survey distributed by email and administered through the survey provider Typeform.

This sample of students is of particular interest as it can be seen as a paradigmatic group of potential STEM entrepreneurs. They have joined an entrepreneurship education program because they already have a business idea to explore and potentially launch, but the setting up of the venture has not started yet (Kelley,

Singer and Herrington, 2016) and they also have technical backgrounds. The main objective of the program is to help students to identify market opportunities, provide tools to develop their ideas and help them to establish a business plan. At the end of the program, the best business ideas are rewarded, and further support is provided to start the business.

The survey was distributed by email to the 1400 applicants, 80% men and 20% women, and the response rate was 10%, making a total final sample of 140 participants. The respondents comprised 82.1% men and 17.9% women, which is representative of the applicant proportions, and the average age was 28. Strict criteria were applied to identify low-quality responses (substantially incomplete; time of completion shorter than 1/3 of the mean; first-click too fast on any of the sections), resulting in a total of 138 valid answers. Questions considered in the analysis was fully completed, with no missing values treatment needed. Geographically, 122 participants were born in Europe (119 in Spain), 14 in America and 3 are from other continents. The major distribution of the participants by areas of knowledge was: Scientific or Technical Degree (94.2%), Social Sciences Degree other than Economics (3.6%) and Other (2.2%). As far as it is a program promoted by a technical university, and since at least one of the team components must have studied there, the strong presence of technical profiles is understandable, which allows the exploration of a STEM sample. At the same time, none of the participants had studied Economics or Business Administration as their main degree, and 5% have a mixed profile with further education in these areas. Some 23.9% had participated in other entrepreneurship education programs before ActuaUPM. Although the program is linked to the university at large, the sample is formed not only of general students but also by post-doctoral students and more mature professional profiles: 72.5% are already graduates (among them 24 are PhD or PhD researchers and 57 hold or were currently studying for, a master's degree) and only 27.5% undergraduates (3 in second course, 7 in third course and 28 in fourth course). 65.9% have work experience. See Table 1 for a summary.

*Table 1. Demographics characteristics of the sample*

|   | No | Percentage |
|---|---|---|
| Sex | 138 | 100% |
|     Men | 114 | 82.6% |
|     Women | 24 | 17.4% |
| Age (mean) | 28,21 | |
| Family main occupations | 138 | 100% |
|     Self-employed | 42 | 30.4% |
|     Private sector | 61 | 44.2% |
|     Public sector | 67 | 48.6% |
|     Other | 6 | 0.04% |
| Participants major | 138 | 100% |
|     Technical or scientific | 130 | 94.2% |
|     Social sciences | 5 | 3.6% |
|     Other | 3 | 2.2% |
| Studies | 138 | 100% |
|     Graduate/Master/PhD | 100 | 72.5% |
|     Undergraduate | 38 | 27.5% |
| Work experience | 138 | 100% |
|     Previous work experience | 91 | 65.9% |
|     Not previous work experience | 47 | 34.1% |
| Entrepreneurship education | 138 | 100% |
|     Previous EEP | 33 | 23.9% |
|     Not previous EEP | 105 | 76.1% |

**Operationalisation**

Dependent variable:

*Entrepreneurship intention* measures the intention of becoming an entrepreneur in the sense of setting up a new business sometime in the future. Entrepreneurship intention is the crucial variable in our analysis and it is very difficult to measure directly. Liñán and Chen, (2009) and Liñán, Rodríguez-Cohard and Rueda-Cantuche (2011) proposed and validated a system based on six items:

*I1- Ready to do anything to be an entrepreneur*

*I2- My professional goal is to be an entrepreneur*

*I3- I will make every effort to start and run my own business*

*I4- I am determined to create a business venture in the future*

*I5-I have very seriously thought in starting a firm*

*I6- I've got the firm intention to start a firm some day*

All items in the questionnaire were measured using a Likert scale, ranging from 1 (totally disagree) to 7 (totally agree). The following section describes the results obtained in relation to these six variables and the analysis of the main components that allows us to obtain the unobserved construct that gathers the maximum information from the items.

Independent variables

The entrepreneurship competences variables were generated following the *EntrComp* model of reference explained above. As far as there are no specific questionnaires exploring this particular framework in previous research, we have defined this block following the questionnaires of previous authors about particular entrepreneurship competences research (DeTienne and Chandler, 2004; Liñán, Rodríguez-Cohard and Rueda-Cantuche, 2011), creating new variables when needed. A first set of 15 variables, aligned with the definitions of the competences proposed, was added to the entrepreneurship intention variables. A small pre-test among PhD researchers and professors was undertaken. Although there were limitations, this pre-test allowed us to measure the average time of response and get feedback about doubts and confusion when answering. Consequently, we modified some variables that included aptitudes that could have different perceptions: (1) communication and leadership skills were separated into two variables, (2) digital skills and legal skills were separated into two variables, (3) networking, team working and problem solving skills were separated into two variables, and (4) "learn by doing" and "learn from mistakes" were also separated. An additional variable referring to "multidisciplinary skills" was added. In this way, the 15 competences described in the framework were captured through 22 variables in the final version of the questionnaire. Before its final launch, the questionnaire was validated in a second pre-test, and there were no problems in understanding the questions and validity of the scales used in this second round. In summary, the theoretical *EntrComp* groups were formed as follows: *Ideas and Opportunities* participants answered five items on 7-point Likert scales, ranging from 1 (no aptitude at all) to 7 (very high aptitude). Six items were built on 7-point Likert scales, ranging from 1 (no aptitude at all) to 7 (very high aptitude) for the *Personal Resources* competences construct. *Specific knowledge* was anchored in three variables built on 7-point Likert scales, ranging from 1 (no aptitude at all) to 7 (very high aptitude). Finally, participants answered five items on 7-point Likert scales, ranging from 1 (no aptitude at all) to 7 (very high aptitude) for the *Into Action* competences.

*Control Variables*

Our analysis also included three control variables: age, having previous entrepreneurship education and self-employed parents. Previous research has shown that age negatively affects entrepreneurial intention (Camelo-Ordaz, Diánez-González and Ruiz-Navarro, 2016). In our research, age was measured as a continuous variable (between 18 and 64). Also the family background has been considered a factor affecting entrepreneurship intention (Carr and Sequeira, 2007). We coded this as a binary variable: 1 if the respondent's parents had been self-employed or were entrepreneurs; 0 if not. Finally, studies about the effect of entrepreneurship education argue that it has a positive impact on intention (Liñán, Rodríguez-Cohard and Rueda-Cantuche, 2011), Souitaris, Zerbinati and Al-Laham, 2007; Sánchez, 2013). The variable values were 1 if the respondent had participated in previous entrepreneurship education programs; 0 otherwise.

**ANALYSIS AND RESULTS**

**Entrepreneurship intention**

Confirming our expectations considering the origin of the sample, participants rate the six items than define *Entrepreneurship Intention* (EI) highly. On a 1-7 scale, measure values are between 4.8 ("*I2-My professional goal is to be an entrepreneur*") and 5.75 ("*I6-I've got the firm intention to start a firm some day*"). In general higher values were obtained for individuals whose parents were entrepreneurs and also for those who had joint previous EEP, although without significant differences.

The ratios of business ventures created in previous editions of this program show that few participants finally become entrepreneurs (Agudo *et al.*, 2014). Agudo et al. (2014) showed that less than 20% of students have historically finished the EEP and less than 10% have finally started up a business. Reasons for this involve the motivation for joining the program, real commitment and the capabilities of the team, so the exploration of the dependent variable can offer interesting insights into the "higher" or "lower" intentions of individuals.

Firstly, we analysed the linear dependency of the six items of the Entrepreneurship Intention construct. The correlation matrix shows high correlations, both positive and significant, between all of them, with values between 0.477 and 0.804. The analysis was completed obtaining Cronbach's alfa showing internal consistency ($\alpha=0.92$), Kaiser-Meyer-Olkin –KMO— as a measure of sampling adequacy (0.883) and Bartlett's Test of Sphericity, highly significant (p value < 0.001), both indicating the dependency between the six variables.

We performed PCA and table 2 displays the loading factors of first component, which explains a high percentage of the variance (74%) and has a high positive correlation with the six variables. We refer to them as *Entrepreneurship Intentions (EI)* and they will be used in the rest of the article as dependant variable. This is considered a suitable measure of the intention to become an entrepreneur, obtained as a combination of the six proposed variables: the higher it is, the greater the intention of individuals to be an entrepreneur. Although all items contribute with similar loadings, I4 and I3 are slightly more important (*determination to create a business in the future and disposition to do any effort to start and run an own business).* Confirming our hypothesis, the intention to become entrepreneurs in this sample was high, with a mean value of 5.34 on a scale of 1 to 7.

*Table 2. EI Component Matrix. Extraction Method: Principal Component Analysis.*

| **Entrepreneurship Intention (EI)** | **Factor 1** |
|---|---|
| I1- Ready to do anything to be an entrepreneur | ,805 |
| I2- My professional goal is to be an entrepreneur | ,861 |
| I3- I will make every effort to start and run my own business | ,900 |
| I4- I am determined to create a business venture in the future | ,910 |
| I5- I have very seriously thought in starting a firm | ,816 |
| I6- I've got the firm intention to start a firm some day | ,866 |

Gender do not significantly affect entrepreneurial intention in this group (t=0.58; p-value=0.954). A direct mean comparison of EI shows similar values for men (5.35) and women (5.33).[4]. There is a widespread idea in the literature according to which the entrepreneurial intention of men is greater than that of women (Guerrero et al., 2016; Kelley et al., 2016; Santos, Roomi and Linan, 2016). In this case, although the sample size (24 women and 114 men) is not large, the means obtained in intention for the two groups are quite similar. Consequently, the results obtained do not confirm the hypothesis of differences between men and women (Hypothesis 1). It is important to note that our sample is not a random sample of the general population but comprises STEM students with an initial predisposition toward entrepreneurship.

**Entrepreneurship competences perception of potential STEM entrepreneurs**

A detailed analysis of this group of potential STEM entrepreneurs shows that the general perception of entrepreneurship competences is quite high, with most values over 4 (the central value). The respondents in general perceive their weakest competences on the *legal and economic know-how,* which was perhaps predictable due to their technical background. The higher rates are obtained in capabilities included in the "*Into Action*" competences area, perceiving themselves as enjoying multidisciplinary profiles that can learn by doing, work in teams, solve problems and take advantage of mistakes.

Although when difference in means is desired for interpretation of the data, Students t-tests analysis are preferred[5], limitations of the data under analysis suggest to use Wilcoxon-Mann-Whitney test for independent sample as a non-parametric method asymptotically more powerful (Fay and Proschan, 2010). In this sense the analysis of differences by gender for these 22 items has been performed using both methods, obtaining similar results (Table 3). In this sample, there are significant differences in four particular capabilities (p-value < 0.05), as explained in the following.

---

[4] Additionally, a discriminant analysis using the six variables and the Wilks' Lambda test indicates no significant differences between the two groups (p-value is 0.820).
[5] Equal variances can be assumed in all items excepting O2 (*Development of creative and purposeful idea*), O9 (*Leadership skills*) and O16 (*Defining priorities and action plans*).

The results suggest that women have a significant lower perception of their ability to *recognise the potential that an idea has for creating value*. This could be aligned with previous research noting that women have a more pessimistic perception of business opportunities (Díaz-Casero, Sanchez and Postigo, 2002; Jack and Anderson, 2002; Camelo-Ordaz, Diánez-González and Ruiz-Navarro, 2016). The difference in the perception of *digital skills*, in which women rate themselves as less capable than men, is also significant. This is especially remarkable in a STEM sample, where all the women participants have a technical background, but previous studies are aligned with this result noting that, although there is no great difference between men and women, regarding digital abilities, women's self-assessed competence is significantly lower (Hargittai, 2006). In the same sense this sample of women perceive themselves to have less *economic and financial know-how* than men, although in both cases it is relatively low. Financial skills are also usually self-rated as lower by women (Brush, 1992; Chaganti, 1986).

By contrast, women's perceptions of their *team working* abilities are higher than men. Team working skills have not been extensively explored in previous research in entrepreneurship competence models but team management is especially valued by women as a pragmatic method of leading teams (A.T. Kearney, 2015).

Consequently, only regarding the *recognition of the potential value of an idea*, the *digital know-how* and the *financial know-how* the capability perceived is higher in men than in women. No significant differences are obtained for the rest of the competences, and in the case of team *working skills*, perceived the capability is higher in women.

*Table 3. Competences perceptions of the sample*

|  | Mean | | | t-test for Equality of Means | | Wilcoxon - Mann - Whitney | |
|---|---|---|---|---|---|---|---|
|  | Male | Fem | Dif | t | p-value | Z | p-value |
| *Ideas and Opportunities (ID)* | | | | | | | |
| O1- Identify opportunities to create value and challenges that need to be met | 5.46 | 5.33 | 0.123 | 0.468 | 0.640 | -0,345 | 0,730 |
| O2- Development of creative and purposeful ideas | 5.84 | 5.83 | 0.009 | 0.030 | 0.976 | -0,608 | 0,543 |
| O3- Visualisation of future scenarios to guide effort and action | 5.60 | 5.83 | -0.237 | -0.995 | 0.322 | -0,836 | 0,403 |
| **O4- Recognise the potential that an idea has for creating value** | **5.44** | **4.79** | **0.647** | **2.518** | **0.013** | **-2,374** | **0,018** |
| O5- Assess the consequences and impact of ideas, opportunities and actions | 5.43 | 5.29 | 0.138 | 0.580 | 0.563 | -0,998 | 0,318 |
| *Personal Resources (PR)* | | | | | | | |
| O6- Identify and assess my individual and group strengths and weaknesses | 5.67 | 5.75 | -0.083 | -0.351 | 0.726 | -0,334 | 0,739 |
| O7- Determination to turn into action my ideas, being resilient under pressure, adversity and temporary failure | 5.71 | 5.79 | -0.081 | -0.284 | 0.777 | -0,900 | 0,368 |
| O8- Making the most of limited resources | 5.82 | 5.88 | -0.059 | -0.227 | 0.821 | -0,270 | 0,787 |
| O9- Leadership skills | 5.70 | 5.54 | 0.160 | 0.515 | 0.611 | -0,106 | 0,916 |
| O-10 Communication skills | 5.58 | 5.88 | -0.296 | -1.065 | 0.289 | -1,304 | 0,192 |
| O11- Multidisciplinary skills | 6.10 | 6.42 | -0.320 | -1.573 | 0.118 | -1,567 | 0,117 |
| *Specific Knowledge (SK)* | | | | | | | |
| **O12- Digital know how** | **5.79** | **5.08** | **0.706** | **2.514** | **0.013** | **-2,456** | **0,014** |
| O13- Legal know how | 3.18 | 3.04 | 0.134 | 0.389 | 0.698 | -0,455 | 0,649 |
| **O14- Financial and economic know how** | **4.11** | **3.46** | **0.656** | **1.974** | **0.050** | **-1,929** | **0,054** |
| *Into Action (IA)* | | | | | | | |
| O15- Development of new products and services | 5.47 | 5.25 | 0.224 | 0.864 | 0.389 | -0,982 | 0,326 |
| O16- Defining priorities and action plans | 5.62 | 5.92 | -0.294 | -1.670 | 0.102 | -1,175 | 0,240 |
| O17- Making decisions dealing with uncertainty, ambiguity and risk | 5.47 | 5.54 | -0.068 | -0.248 | 0.804 | -0,428 | 0,669 |
| O17- Networking skills and making professional contacts | 4.92 | 4.79 | 0.129 | 0.373 | 0.710 | -0,441 | 0,659 |
| **O19- Team working** | **5.90** | **6.42** | **-0.513** | **-2.177** | **0.031** | **-2,372** | **0,018** |
| O20- Problem solving skills | 5.93 | 6.29 | -0.362 | -1.727 | 0.086 | -1,885 | 0,059 |
| O21- Learn by doing | 6.37 | 6.46 | -0.090 | -0.496 | 0.620 | -0,367 | 0,713 |
| O22- Learn from mistakes | 6.32 | 6.33 | -0.018 | -0.088 | 0.930 | -0,115 | 0,909 |

**Identification of dimensions of competences**

The first analysis performed aimed to test whether the *EntrComp* competences theoretical constructs was statistically valid[6] and the confirmatory analysis did not finally support the theoretical four constructs grouping proposed by the *EntrComp* (Estimator ML, Model Fit Statistic 393.538, Degrees of freedom 203, P-value Chi-square 0.000).

Taking this into account, from the theoretical approach based on the *EntrComp* Framework, factor analysis was used to determine the smallest number of factors to best represent the inter-relationships among the set of 22 self-reported competencies, and to identify the competencies that loaded onto the key factors. The suitability of the data for the PCA was established by various means. The Cronbach's alpha coefficient was calculated; with a value of 0.88 this confirmed the reliability of the scale within the sample. The alpha coefficient is not improved by removing any of the items, so the factor analysis included the whole set of 22 variables.

Both Kaiser-Meyer-Olkin (KMO) and Bartlett's test of sphericity were conducted to measure sampling adequacy. The KMO value was 0.83, which is greater than the recommended value of 0.6 (Kaiser, 1974). Bartlett's test was statistically significant at the p 0.00 level (Bartlett, 1954).

Factor analysis was executed by extracting the number of factors with eigenvalues >1; this resulted in the identification of six factors for a satisfactory model fit (chi square statistic is 126.4 on 114 degrees of freedom; p-value is 0.201).

---

[6] The theoretical grouping was maintained and a principal components analysis (PCA) was carried out in each block of items. The scores obtained for the first components in each analysis were chosen as an indirect measure of the unobservable constructor. Principal components analysis is used for dimension reduction and the loading factors of variables are showed in Table 4. SPSS software was used for the analysis. The analysis obtains an index (construct) for each block that is a linear combination of the items that define it. This indicator extracts the maximum information from the group of associated items.

We used a Cronbach's test to test the reliability and validity of the latent variables (constructs), obtaining satisfactory internal consistency for all the constructs proposed (EI 0.92; ID 0.78; PR 0.70; SK 0.70; IA 0.74). The Kaiser-Meyer-Olkin measure of sampling adequacy (KMO), which indicates the proportion of variance that might be caused by underlying factors, was above 0.55 and around 0.80 for most factor analysis (*EI* KMO=0.88; *ID* KMO=0.80; *PR* KMO=0.64; *SK* KMO =0.57; *IA* KMO=0.67). Bartlett's tests of sphericity are highly significant in all cases, which indicated that a factor analysis may be useful. There were strong correlations in Entrepreneurship Intention (EI) in all blocks of competences, which suggests multicollinearity issues. The confirmatory analysis did not finally support the theoretical four constructs grouping proposed by the *EntrComp* (Estimator ML, Model Fit Statistic 393.538, Degrees of freedom 203, P-value Chi-square 0.000).

Next, the factors were rotated using Varimax with Kaiser normalisation to avoid multicollinearity effects among the variables. Item 22, learn from mistakes, load onto two components, which is not surprising because the own *EntrComp* framework stands they are interconnected and previous studies looking for competence clustering obtained similar results regarding the permeability of boundaries (Mitchelmore and Rowley, 2013). Component 2 gathers those competences related to commitment and component 6 includes item 22 isolated. We used a Cronbach's test to evaluate the reliability and validity of the components and the Cronbach's alpha of factor 2 with item 22 is not worse than without it, then we consider this variable loads in Factor 2 and discard the component 6. Satisfactory internal consistency is obtained for all the components.

Table 4 lists the eigenvalues associated with these five actors, and the variance in self-reported competencies explained. The selected five components explain a total of 57.372 per cent of the variance.; it shows meaningful strong loadings for each of the five components. Four competences do not load significantly on any of the components: *Development of creative and purposeful ideas (O2)*, *development of new products and services (O15)*, *networking skills and making professional contacts (O18)* and *problem solving skills (O20)*.

*Table 4. Result of factor analysis for entrepreneurship competences*

| Factor | Eigen value | %variance | α | Variable name | Factor loading | Operational definition |
|---|---|---|---|---|---|---|
| 1 | 6.683 | 30.378 | 0.71 | O4- Recognise the potential that an idea has for creating value | 0.751 | Opportunity and ideas (ID) |
| | | | | O3- Visualisation of future scenarios to guide effort and action | 0.711 | |
| | | | | O1- Identify opportunities to create value and challenges that need to be met | 0.668 | |
| | | | | O5- Assess the consequences and impact of ideas, opportunities and actions | 0.603 | |
| 2 | 1.9 | 8.638 | 0.71 | O21- Learn by doing | 0.686 | Commitment competences (CC) |
| | | | | O8- Making the most of limited resources | 0.658 | |
| | | | | O12- Digital know how | 0.579 | |
| | | | | O7- Determination to turn into action my ideas, being resilient under pressure, adversity and temporary failure | 0.564 | |
| | | | | O22- Learn from mistakes | 0.533 | |
| | | | | O11- Multidisciplinary skills | 0.507 | |
| 3 | 1.571 | 7.142 | 0.65 | O16- Defining priorities and action plans | 0.787 | Decision making (DM) |
| | | | | O17- Making decisions dealing with uncertainty, ambiguity and risk | 0.669 | |
| | | | | O6- Identify and assess my individual and group strengths and weaknesses | 0.613 | |
| 4 | 1.268 | 5.765 | 0.67 | O10- Communication skills | 0.836 | Organising competences (OC) |
| | | | | O19- Team working | 0.625 | |
| | | | | O9- Leadership skills | 0.542 | |
| 5 | 1.199 | 5.449 | 0.83 | O14- Financial and economic know how | 0.867 | Specific knowledge (SK) |
| | | | | O13- Legal know how | 0.849 | |

The model results in five factors, which offer a different categorisation from the *EntrComp* theory and suggest an aggrupation aligned with Man et al. (2002), with exception of the relationship competences related to the context such as networking and professional contacts, which does not significantly load in any of the factors resulting in our model.

Factor 1 – *Opportunity and ideas (ID)*. Four items cluster to form the first factor. This factor includes most of the competences linked to the first group of *EntrComp*, referring to the ability to recognise opportunities, value ideas, forecast future scenarios and evaluate consequences. From the *EntrComp* theory framework, the one that does not load into this factor is the perception of the ability to *develop creative and purposeful idea*, which does not significantly load in any of the components.

Factor 2 – *Commitment competences (CC)*. Six items cluster to form the second factor. This includes competences related to moving ahead with the business: determination to turn ideas into action, being resilient under pressure and temporary failure, making the most of limited resources, learning by doing and learning from mistakes. The perception of having multidisciplinary skills and digital know how also contributes to his factor.

Factor 3 – *Decision making (DM)*. Three items cluster to form the third factor: defining priorities and action plans, making decisions with uncertainty and risk and the identification and assessment of the own individual and group strengths and weaknesses.

Factor 4 – *Organising competences (OC)*. Three items cluster to form the fourth factor, related to team management: communication skills, leadership and team working.

Factor 5 – *Specific knowledge (SK)*. Two items cluster to form the fifth factor: and they correspond to the legal and financial and economic knowledge.

**Regression analysis**

Once the factors were identified, we used a multiple regression model to study the relationship between the competence and skill factors in *Entrepreneurship Intention (EI)*.

The model of competences proposed has a R-square=.284. Hypothesis 2 is validated for ID, CC, DM and OC competences, which significantly affect the intention. The relationships between SK and EI is not validated in this sample.

When considering gender in the model, results reveal that the relation between competences and intentions in this sample is not moderated by gender gender and therefore hypothesis 3 is not validated. None of the other control variables reached the level of significance. Table 5 summarises the parameters obtained.

*Table 5. Regression components of the sample towards intention*

| Predictor variables | EI | Sig | EI | Sig | EI | Sig | EI | Sig | EI | Sig | EI | Sig | EI | Sig | EI | Sig | EI | Sig | EI | Sig | EI | Sig | EI | Sig |
|---|---|---|---|---|---|---|---|---|---|---|---|---|---|---|---|---|---|---|---|---|---|---|---|---|
| ID (Component 1) | | | | | | | | | | | .354** | .000 | .353** | .000 | .362** | .000 | .362** | .000 | .358** | .000 | 0,372** | .000 | 0,339** | .000 |
| CC (Component 2) | | | | | | | | | | | .300** | .000 | .300** | .000 | .303** | .000 | .305** | .000 | .301** | .000 | 0,309** | .000 | 0,325** | .000 |
| DM (Component 3) | | | | | | | | | | | .165* | .027 | .166* | .028 | .168* | .026 | .166* | .026 | .154* | .044 | 0.158* | 0,04 | 0,162* | 0,06 |
| OC (Component 4) | | | | | | | | | | | .170* | .023 | .173* | .024 | -.172* | .022 | .179* | .017 | .181* | .018 | 0.191* | 0,02 | 0,180* | 0,04 |
| SK (Component 5) | | | | | | | | | | | .057 | .440 | .055 | .472 | .060 | .423 | .054 | .474 | .051 | .492 | 0.049 | 0,53 | 0,043 | 0,64 |
| | | | | | | | | | | | | | | | | | | | | | | | | |
| ID*GENDER | | | | | | | | | | | | | | | | | | | | | | | 0,103 | 0,25 |
| CC*GENDER | | | | | | | | | | | | | | | | | | | | | | | -0,057 | 0,54 |
| DM*GENDER | | | | | | | | | | | | | | | | | | | | | | | -0,049 | 0,62 |
| OC*GENDER | | | | | | | | | | | | | | | | | | | | | | | 0,098 | 0,37 |
| SK*GENDER | | | | | | | | | | | | | | | | | | | | | | | 0,025 | 0,8 |
| | | | | | | | | | | | | | | | | | | | | | | | | |
| Gender | .005 | .954 | | | | | | | -.006 | .946 | | | -.145 | .885 | | | | | | | -0.003 | .970 | -.13 | .876 |
| Previous EEP | | | 0.72 | .402 | | | | | .077 | .378 | | | | | -0.043 | .572 | | | | | -0.035 | .656 | -0,034 | 0,67 |
| Age | | | | | .023 | .790 | | | .21 | .808 | | | | | | | -0.066 | .387 | | | -0.065 | .397 | -0,064 | 0,45 |
| Family entrepreneur | | | | | | | .042 | .622 | .594 | .553 | | | | | | | | | .067 | .385 | 0.063 | .419 | 0,058 | 0,47 |
| R2 | .005 | | 0.72 | | .023 | | .042 | | .091 | | .524 | | .524 | | .526 | | .528 | | .528 | | .533 | | .544 | |
| Adjusted R2 | .000 | | 0.05 | | .001 | | .002 | | .008 | | .275 | | .275 | | .277 | | .279 | | .279 | | .284 | | .296 | |

**. Significant at the 0.01 level (2-tailed). *. Significant at the 0.05 level (2-tailed).

Competences loading in *Opportunity an Ideas (ID)*, regarding the *ability to identify opportunities and value of the potential of ideas* results the most influential competence towards the entrepreneurship intention (β=0.339; t=3.983;p-value=0.000). Opportunity recognition capacity in the framework of entrepreneurship education has been explored in recent years, included in previous entrepreneurship competences frameworks (Man, Lau and Chan, 2002; Mitchelmore and Rowley, 2010, 2013) and found to significantly entrepreneurship intention (Dimov, 2007; Santos, Marques and Ferreira, 2018). This ability is considered one of the most important attributes for successful entrepreneurs (Dimov, 2007) and, despite the creativeness or motivation of the potential entrepreneur, some scholars note that without identifying opportunities to target, entrepreneurial activity does not take place (Short *et al.*, 2010). Our results show that this ability significantly affects the entrepreneurship intentions of this group of STEM students. The second group of competences that results more related to a higher entrepreneurship intention is related to *Commitment Competences (CM) (*β=0.325; t=3.803; p-value=0.000). A positive relation is also found between a higher perception of abilities related to DM, related to *making decision (*β=0.162; t=1.895; p-value=0.060) and OC, regarding the *organization of teams (*β=0.180; t=2.050; p-value=0.042). Action-oriented competences have been considered key abilities in previous models (Boyles and College, 2012). On the other side, the effect of higher levels of specific knowledge, such as economic, financial or legal skills, towards intention has not been validated in this sample. This scenario is aligned with trends in entrepreneurship education models, in which the development of soft skills plays a higher role than business or management contents (Haase and Lautenschläger, 2011) and suggests a need to reinforce methods that contribute to developing personal skills and entrepreneurial mindsets rather than specific management knowledge (Costa *et al.*, 2018).

**Conclusions**

This paper contributes to an understanding of the entrepreneurship competences held by a group of potential STEM entrepreneurs by identifying differences by gender and exploring which ones are more closely related to higher entrepreneurship intention, which can contribute to orienting education initiatives.

The literature review conducted shows the difficulties of defining a comprehensive framework of what has been called entrepreneurship competences and, as a reference for the analysis, this paper draws on the *EntrComp* theoretical framework proposed as a common guide to inspire entrepreneurship education by the European Commission.

In the first part of the analysis, we have explored the self-perception of competences. The results from the analysis show that female participants feel less competent on three variables: the *recognition of the potential value of an idea*, the *digital know-how* and the *financial know-how*, but they rate the *team working skills* higher than male participants. Despite previous studies usually point out that women rate their entrepreneurship competences lower than men, in this STEM sample there are not significant differences by gender excepting on those four variables. Considering the whole sample, these STEM potential entrepreneurs perceived themselves as multidisciplinary profiles able to learn by doing, work in teams, solve problems and take advantage of mistakes.

Next, the analysis of the entrepreneurship intention confirm a all the individuals share the common interest on becoming entrepreneurs and reveal that entrepreneurship intention between both female and male profiles is not significantly different in this STEM case study, despite the widespread hypothesis that entrepreneurship intention is lower in women than in men.

Finally, we have used regression analysis considering gender as a moderating factor in order to explore which competences were more closely related to a higher entrepreneurship intention. Factor analysis conducted suggest a clustering of competences that does not follow the theoretical groups proposed by the *EntrComp*. The *EntrComp* proposal offers a grouping conceptualized on the basis of a process that takes ideas into action (Opportunities & Ideas, Resources and Into Action). Nevertheless, a purpose-behaviour classification aligned with previous frameworks such as Man et al. (2002) fits a better model and the resulting analysis considers five areas: *Opportunity and Ideas competences*, *Commitment competences*, *Decision making competences, Organising competences and Specific knowledge.*

The competences area of *opportunities and ideas (ID)*, regarding the ability to identify opportunities and value of the potential of ideas, results the most influential towards the entrepreneurship intention. The dimensions including those competences related to the *commitment* to work on the business (CC), to the *decision-making* process (DM) and to the *organization (OC)*, also result significantly related to a higher entrepreneurship intention.

This relation between competences perceived and entrepreneurship perception does not result being moderated by gender in this STEM sample, despite previous research finds the lack of competence to be more important for women than men, in the sense of having the necessary knowledge, skills, and experience to be an entrepreneur

Overall, as a contribution from the analysis, the inclusion of methods that promote competences oriented to the *identification of opportunities*, *commitment*, *decision-making* and *organization* can contribute entrepreneurship education programs as a tool that foster both STEM women and men entrepreneurship intention. Our results do not, however, identify specific orientations that effect female more than male participants. Incorporating methods that teach the use of prototypes and experimentation in order to identify an opportunity into entrepreneurship education programs could positively affect STEM entrepreneurship intention. Experiential learning, proposing solutions to problems, learning by doing and engaging in real-life situations can be promoted by the incorporation of entrepreneurship experiences in classes, and real environment observation to identify opportunities (Costa *et al.*, 2018)..

On the other hand, the effect of higher levels of specific knowledge on intention, such as financial or legal skills, has not been validated in this sample. This scenario is aligned with entrepreneurship education model trends, in which the development of soft skills plays a higher role than business or management contents (Haase and Lautenschläger, 2011) and suggests reinforcing methods that contribute to develop personal skills and entrepreneurial mindset more than specific management knowledge (Costa *et al.*, 2018).

**LIMITATIONS AND FURTHER RESEARCH**

Given the specific focus of this paper, and although the availability of data for potential STEM women entrepreneurs is limited and difficult to gather, the results presented encourage further exploration of the effects of entrepreneurial competences in other programs and target groups to test the relationship not only with intention, but also with business performance, once the stand-up phase has been accomplished. More sophisticated data analysis methods could offer greater insight into complex factors that affect relationships.

The current debate lays down over the competences that entrepreneurship education should pursue. From the authors' perspective, an understanding of which abilities are better related to the entrepreneurship intention of potential entrepreneurs is the first step, and should only be considered as a departing point from which to define and test methods and practices that could configure comprehensive entrepreneurship education programs, adapted to promote specific competences that may have an impact on entrepreneurial intention and subsequent behaviour. Finally, observation of the evolution of these competences through any entrepreneurship program and the intention to start a business could help to evaluate the effectiveness of the methodologies and contents taught and contribute to further research into the impact of entrepreneurship education.